\documentclass[eqsecnum,showpacs,aps]{revtex4}
\usepackage{epsfig}

\newcommand{\beq}{\begin{equation}}
\newcommand{\eeq}{\end{equation}}
\newcommand{\beqa}{\begin{eqnarray}}
\newcommand{\eeqa}{\end{eqnarray}}
\def\ket#1{|\,#1\,\rangle}
\def\bra#1{\langle\, #1\,|}

\def\proj#1#2{\ket{#1}\bra{#2}}

\def\ol#1{\overline{#1}}
\def\kpsi{\ket{\psi}}

\def\opone{\leavevmode\hbox{\small1\kern-3.8pt\normalsize1}}

\begin{document}
\title{Asymmetric quantum telecloning of $d$-level systems and
  broadcasting of entanglement to different locations  using
  the "many-to-many" communication protocol}

\author{Iulia Ghiu$^{1,2}$} \thanks{Email: iughiu@barutu.fizica.unibuc.ro} 
\affiliation{$^1$ Department of Microelectronics and Information Technology, \\
Royal Institute of Technology (KTH), Electrum 229, 164 40 Kista, Sweden
\\
$^{2}$ Department of Physics, Section of Statistical Physics and Quantum Mechanics, \\ University of Bucharest, PO Box MG-11, R-76900, Bucharest-M\u{a}gurele, Romania}

\date{\today}
%\draft

\begin{abstract}
We propose a generalization of quantum
teleportation: the so-called many-to-many quantum
communication of the information of a $d$-level system from $N$ spatially
separated senders to
$M>N$ receivers situated at different locations.
We extend the concept of asymmetric telecloning from qubits to
$d$-dimensional systems. We investigate the broadcasting of entanglement by
using local $1 \to 2$ optimal universal asymmetric Pauli machines and
show that the maximal fidelities of the two final entangled states are obtained when
symmetric machines are applied. Cloning of entanglement is studied
using a nonlocal optimal universal asymmetric cloning machine and we
show that the symmetric machine optimally copies the entanglement.
The "many-to-many" teleportation scheme is applied in order to distribute entanglement shared between two observers to two pairs of
spatially separated observers.
 
\end{abstract}

\maketitle

\section{Introduction}
Multiparty entanglement plays a crucial role in most applications of the quantum information processing, e.g., quantum teleportation \cite{Bennett}, quantum key distribution required in cryptography \cite{Ekert,Hillary}, superdense coding \cite{Bennett2}, entanglement swapping \cite{Bose}, and quantum computation \cite{Shor,Grover}.

In the original work of Bennett {\it et al.} \cite{Bennett}, quantum
teleportation of the information contained in a state of an unknown $d$-level system was
introduced. In this scheme, an observer Alice transmits the
information of the particle to another observer Bob, with perfect
fidelity, by using a maximally entangled state. 
Recently, a more general scheme has been proposed: the so-called
one-to-many quantum transmission of the information of a $d$-level
particle from one sender to $M$ receivers located at different places \cite{Murao2}. The task is achieved if Alice performs a measurement in the
generalized Bell basis and sends the outcome to the receivers, who
will apply a local ``recovery unitary operation'' (LRUO) in order to obtain the initial state. 

In this paper, we propose
a different generalization of quantum teleportation, namely, the {\it
  many-to-many} quantum transmission of the information of a $d$-level
system from $N$ senders spatially separated to $M$ receivers (with
$M>N$). In this scheme, the unknown state to be teleported is an
entangled $N$-particle state shared by $N$ observers. The quantum channel
used for broadcasting the information 
between the senders and receivers is 
a maximally entangled $(N+M)$-party state.
 Each sender has to perform a Bell-type measurement and announce the
 outcome to the receivers, who perform a LRUO, obtaining the
 information of the initial unknown state.

Cloning (copying of a system) is a process
that shows the significant differences between classical and quantum
information processing. In contrast to the classical case, when one can
generate as many copies of a system as one wishes,
 Wootters and Zurek have shown that no 
machine exists that can produce two perfect copies of an unknown quantum
state \cite{Wootters}; this statement was called the {\it no-cloning
theorem}. Since perfect cloning is not possible, a new question has
arisen, namely, how well can we copy an unknown quantum state? 
Cloning machines can be divided into symmetric and asymmetric machines.
 A universal $1 \to 2$
symmetric quantum cloner is defined as a machine that transforms an
input qubit $\kpsi $ into two identical imperfect copies (clones), characterized by
the density operator \cite{Buzek,Bruss}
\beq
 \rho=s\proj{\psi}{\psi}+\frac{1-s}{2}I.\eeq  
The
cloner is optimal if the fidelity of the final states is maximal.
In the last few years, much progress has been made in analyzing the optimal
cloning of different systems: a $1 \to 2$ universal cloning machine for qubits \cite{Bruss}, a $1 \to 2$ symmetric cloner for $d$-level
systems \cite{Buzek3}, an $N \to M$ symmetric cloner for qubits (a machine that
takes as input $N$ identical qubits and generates at the output $M>N$ copies)
\cite{Gisin,Bruss2}, and an $N \to M$ symmetric cloner for $d$-level systems
\cite{Werner,Keyl,Zanardi,Fan}.
While a symmetric machine produces identical output states, the
$1 \to 2$ asymmetric cloner of qubits proposed by Cerf
 generates two
output states emerging from two different Pauli channels
\cite{Cerf0}. The asymmetric cloning machine has been extended to higher
dimensions (a Heisenberg cloning machine) also by Cerf
\cite{Cerf1}. Experimental implementations of both quantum
teleportation \cite{Bouwmeester}, and, recently, quantum cloning of
photons \cite{Lamas,Fasel} have been carried out. 

An interesting application of quantum cloning is the broadcasting of
 entanglement proposed by  Bu$\check z$ek {\it et al.} \cite{Buzek2}. In this
process, the entanglement originally shared by two observers is
broadcast into two identical entangled states by using a local $1 \to
2$ optimal universal symmetric cloning machine. In Ref. \cite{SB},
this process has been investigated  with the help of a local universal
symmetric cloning machine (not necessary optimal) and it has been shown
that the entanglement is optimally split when optimal symmetric
cloners are used. 

In this paper we analyze the broadcasting of
entanglement by considering local optimal
  universal {\it asymmetric} cloning machines and prove that the two output
  entangled states are different, but have the same fidelity. We show
  that the inseparability is optimally broadcast when symmetric
  cloners are applied. In
  Ref. \cite{Buzek3}  cloning of
  entanglement using nonlocal symmetric cloners was proposed. We investigate the
  cloning of entanglement by applying a nonlocal optimal
  universal asymmetric cloning machine and also find that
  symmetric cloners generate the best copies. 

In Ref. \cite{Murao1}, Murao {\it et al.} proposed a quantum scheme
called telecloning that combines quantum teleportation and optimal
cloning. In this protocol,
an observer Alice has to send $M$ identical copies of an unknown qubit to $M$
spatially separated observers. Due to the no-cloning theorem
\cite{Wootters}, an unknown quantum state cannot be copied
faithfully; therefore Alice can send only imperfect copies to the $M$
receivers. The task is achieved with the help of a multiparty
quantum information distribution channel shared between the senders
and receivers.
  D\"{u}r and Cirac presented a generalization
 of quantum telecloning, where Alice holds $N$ identical qubits and has
 to send the copies to $M$
 observers ($M>N$) situated in different locations. Instead of performing a
 Bell measurement, Alice uses a more generalized positive operator
 valued measure  \cite{Dur}. By applying the one-to-many teleportation
 protocol, Murao {\it et al.} gave further generalizations of
 telecloning: the $1 \to M$ symmetric telecloning of a $d$-level system and the
 $1 \to 2$ asymmetric telecloning for qubits
  \cite{Murao2}.

In this paper we present a generalization of asymmetric telecloning from
qubits to $d$-level systems. We also propose a scheme called {\it telebroadcasting}, which performs the
broadcasting of entanglement to different locations using the many-to-many communication protocol. A summary of earlier papers and new results of the
present work on teleportation, cloning, telecloning, broadcasting of
entanglement, and telebroadcasting is given in Table I.

The paper is organized as follows. In Sec. II we start by reviewing the standard
teleportation scheme for $d$-level systems and its generalization, one-to-many
teleportation. Then we propose a generalization called the
"many-to-many" communication, a procedure that involves $N\ge 1$
senders situated at different locations. In Sec. III we give a summary
 of the asymmetric Heisenberg quantum cloners for $d$-level systems,
analyzing the special case of optimal universal asymmetric Heisenberg
machines using a different, but equivalent, description to the one given in
Ref. \cite{Cerf1}. The extension of asymmetric telecloning from qubits
to $d$-level systems is presented in 
Sec. III C. Broadcasting of inseparability using local and nonlocal optimal
  universal asymmetric cloning machines is investigated in Sec. IV. 
Applying the generalized many-to-many teleportation protocol, we
propose in Sec. V $1 \to 2$ telebroadcasting, a process that combines
teleportation and broadcasting of an arbitrary bipartite two-level
entangled state to two pairs of spatially separated observers.  Our
conclusions are summarized in Sec. VI.

\newpage
\begin{table}
{\bf Teleportation} 

\begin{tabular}[t]{|c|c|c|c|}
{\bf Dimension}   & {\bf Number of senders} & {\bf Number of receivers} &{\bf Authors}  \\ \hline
$d$   & 1 &1  & Bennett {\it et al.} \cite{Bennett}      \\ \hline
$d$    & 1  & $M$ & Murao {\it et al.} \cite{Murao2}     \\ \hline
$d$   & $N$  & $M>N$  & {\it present work}     
\end{tabular}

{\bf Symmetric cloning}

\begin{tabular}[t]{|c|c|c|c|}
{\bf Dimension}   & {\bf Number of copies of the initial system} &{\bf Number of final
clones }& {\bf Authors}  \\ \hline
2   & 1& 2 &  Bu$\check z$ek and Hillery \cite{Buzek}, Bru{\ss}  {\it et al.}  \cite{Bruss}  \\ \hline
$d$   & 1 &2  & Bu$\check z$ek and Hillery \cite{Buzek3}      \\ \hline
2    & $N$  & $M>N$ & Gisin and Massar \cite{Gisin}, Bru{\ss} {\it et al.}  \cite{Bruss2}    \\ \hline
$d$   & $N$  & $M>N$  & Werner\cite{Werner}, Keyl and Werner \cite{Keyl},\\
&&&Zanardi \cite{Zanardi}, Fan {\it et al.} \cite{Fan}     
\end{tabular}

{\bf Asymmetric cloning}

\begin{tabular}[t]{|c|c|}
{\bf Dimension}    & {\bf Authors}  \\ \hline
2   &   Cerf \cite{Cerf0}   \\ \hline
$d$   &  Cerf \cite{Cerf1}
\end{tabular}

{\bf Broadcasting of entanglement}

\begin{tabular}[t]{|c|c|}
{\bf Cloning machine}    & {\bf Authors}  \\ \hline
Local Symmetric   &Bu$\check z$ek {\it et al.} \cite{Buzek2}   \\ \hline
Nonlocal Symmetric    & Bu$\check z$ek {\it et al.} and Hillery    \\ \hline
Local Asymmetric    & {\it present work}      \\ \hline
Nonlocal Asymmetric    & {\it present work}     
\end{tabular}

{\bf Telecloning} 

\begin{tabular}[t]{|c|c|c|c|c|}
{\bf Telecloning}&{\bf Dimension}   & {\bf Number of copies of the
  initial system}& {\bf Number of receivers} & {\bf Authors}  \\ \hline
Symmetric&2   & 1& $M$ & Murao {\it et al.}  \cite{Murao1}  \\ \hline
Symmetric&2  & $N$ &$M>N$  & D\"{u}r and Cirac  \cite{Dur}    \\ \hline
Symmetric&$d$    & 1  & $M$ & Murao {\it et al.}  \cite{Murao2}   \\ \hline
Asymmetric&2   & 1  & 2  & Murao {\it et al.} \cite{Murao2}   \\ \hline  
Asymmetric&$d$   & 1  & 2  & {\it present work}
\end{tabular}

{\bf Telebroadcasting}

\begin{tabular}[t]{|c|c|c|c|}
{\bf Telebroadcasting}&{\bf Number of senders}&{\bf Number of
  receivers}    & {\bf Authors}  \\ \hline
Asymmetric&1$\times $2   &2$\times $2&   {\it present work}
\end{tabular}
\caption[]{Summary of earlier work and the present results on
  teleportation, cloning, telecloning, broadcasting of entanglement, and telebroadcasting.}
\label{Table1}
\end{table}

%-----------------------------------------------------------------------------------------
%-----------------------------------------------------------------------------------------

\section{A generalization of quantum teleportation}

\subsection{Preliminaries}
Let us start by giving a summary of the original teleportation protocol and its
generalization $-$ the one-to-many scheme for sending quantum information.
In the first quantum teleportation scheme proposed by Bennett {\it et
al.} \cite{Bennett}, an unknown state of a $d$-level system is faithfully
transmitted from one observer Alice to another observer Bob, with
the help of quantum and classical channels.
The initial unknown state we wish to teleport is
\beq
\kpsi=\sum_{k=0}^{d-1}\alpha_k\ket{k}_A,
\eeq
where  $\sum_{k=0}^{d-1}|\alpha_k|^2=1$, and $\{\ket{k}\}$ is the
computational basis.
In order to perform the teleportation, Alice and Bob must share a
quantum channel, which is a
maximally entangled state:
\beq
\ket{\xi}=\frac{1}{\sqrt d}\sum_{j=0}^{d-1}\ket{j}_A\ket{j}_B.
\eeq
The complete state of the system can be written as
\beq
\kpsi\ket{\xi}=\frac{1}{d}\sum_{m=0}^{d-1}\sum_{n=0}^{d-1}\ket{\Phi_{m,n}}\sum_{k=0}^{d-1}\mbox{exp}\left(
  -\frac{2\pi ikn}{d}\right) \alpha_k\ket{\ol{k+m}},
\eeq
where $\ol{k+m}=k+m$ modulo $d$, and
\beq \label{fimn}
\ket{\Phi_{m,n}}=\frac{1}{\sqrt d}\sum_{k=0}^{d-1}\mbox{exp}\left(\frac{2\pi ikn}{d}\right)\ket{k}\ket{\ol{k+m}},
\eeq
is the generalized Bell basis \cite{Bennett,Cerf1}.
Alice performs a Bell-type measurement on her particles and sends
the result to Bob (classical communication). Suppose the outcome of
Alice's measurement is $\ket{\Phi_{m,n}}$. Bob has to apply the
unitary operator \cite{Bennett}
\beq
V_{m;n}=\sum_{j=0}^{d-1}\mbox{exp}\left( \frac{2\pi ijn}{d}\right) \ket{j}\bra{\ol{j+m}}
\eeq
on his particle in order to recover the initial state. This protocol
was called the ``one-to-one'' quantum communication of
teleportation in Ref. \cite{Murao2}, because the
information has been transmitted from one sender to one receiver.

%-----------------------------------------------------------------------------------------

Let us now review the one-to-many teleportation protocol proposed by
Murao {\it et al.} \cite{Murao2}, where the information  is distributed from
one sender Alice to many distant
receivers $B_1$, $B_2$,..., $B_{N_o}$, using multiparty entanglement. Alice's system is an
$N_i$-particle state, which encodes the information of a
$d$-level system:
\beq \label{1m}
\kpsi=\sum_{k=0}^{d-1}\alpha_k\ket{\psi_k}_A,
\eeq
with $\sum_{k=0}^{d-1}|\alpha_k|^2=1$, and $\{\ket{\psi_k}_A\}$
represents a basis in the $d$-dimensional space.
The quantum channel used in this scheme is a maximally entangled state
of Alice's $N_p$ particles and the receivers' $N_o$ particles:
\beq
\ket{\xi}=\frac{1}{\sqrt d}\sum_{j=0}^{d-1}\ket{\pi _j}_A\ket{\phi
  _j}_{B_1B_2...B_{N_0}}.
\eeq
We shall follow the same procedure as before:
Alice measures her particles in the generalized $(N_i+N_p)$-particle
Bell basis:
\beq \label{GenBell}
\ket{\Phi_{m,n}}=\frac{1}{\sqrt d}\sum_{k=0}^{d-1}\mbox{exp}\left(\frac{2\pi
    ikn}{d}\right) \ket{\psi_k}\ket{\pi_{\ol{k+m}}},
\eeq
and then she announces the outcome.

The recovery unitary operation (RUO) \cite{Murao2} which is implemented by
the receivers has to be a local one, and satisfies the condition
\beq  \label{onetomany}
V_{m;n}={\cal V}_{B_1}\otimes {\cal V}_{B_2}\otimes ...\otimes {\cal V}_{B_{N_o}}=\sum_{j=0}^{d-1}\mbox{exp}\left(\frac{2\pi ijn}{d}\right)\ket{\phi_j}\bra{\phi_{\ol{j+m}}}.
\eeq
Hence, the receivers will share a state that contains the information
of the initial state of Eq. (\ref{1m}):
\beq \label{otm}
\ket{\phi}=\sum_{j=0}^{d-1}\alpha_j\ket{\phi_j}_{B_1B_2...B_{N_o}}.
\eeq

%-----------------------------------------------------------------------------------------

\subsection{A generalization of quantum teleportation: The many-to-many communication protocol}

In this subsection we present a generalization of teleportation called the ``many-to-many'' communication of a $d$-level system.
Assume that an entangled state, which contains the information of a $d$-level system, is shared by $N$ observers $A_1, A_2,..., A_N$, spatially separated:
\beq \label{initial}
\kpsi=\sum_{k=0}^{d-1}\alpha_k\ket{\psi_k}_{A_1}\ket{\psi_k}_{A_2}...\ket{\psi_k}_{A_N},
\eeq
where $\sum_{k=0}^{d-1}|\alpha_k|^2=1$, and $\{\ket{\psi_k}_{A_j}\}$ represents a basis in the $d$-dimensional space of the $j$th sender.
The task is to transmit the quantum information of this state to $M$
receivers (with $M>N$) situated at different locations; therefore we propose a generalization of the teleportation protocol introduced in Ref. \cite{Murao2} from one sender to $N$ senders.

Let us consider the following scenario. The senders perform local operations on their
particles and $M-N$ ancillas in order to encode the information of the
initial $N$-particle state into the final $M$-particle state. Then, the
senders send the information of the $M$-particle state to the receivers by using the
standard teleportation scheme \cite{Bennett} that requires an
entanglement 
\beq
E_1=M\mbox{log}_2d
\eeq
 between the senders and receivers,
and $M$ Bell-type measurements. A similar scenario was analyzed in Ref. \cite{Murao2}, where the initial
$N$-particle state belongs to Alice.

Now we investigate a second scenario where the quantum channel used is a maximally entangled $(N+M)$-particle state shared between senders and receivers:
\beq
\ket{\xi}=\frac{1}{\sqrt d}\sum_{j=0}^{d-1}\ket{\pi _j}_{A'_1}\ket{\pi _j}_{A'_2}...\ket{\pi _j}_{A'_N}\ket{\phi _j}_{B_1B_2...B_M},
\eeq
where we have denoted by $B$ the particles that belong to the receivers. The states $\{\ket{\pi _j}_{A'_i}\}$ represent a $d$-dimensional basis for the $i$th sender. 
In this case
the entanglement of the bipartition senders$-$receivers of the channel is $E_2=\mbox{log}_2d<E_1$.
 The state of the whole system is
\beqa
&&\kpsi\ket{\xi}=\frac{1}{\sqrt d}\sum_{k=0}^{d-1}\alpha_k\sum_{j=0}^{d-1}\ket{\psi_k}_{A_1}\ket{\pi_j}_{A'_1}\ket{\psi_k}_{A_2}\ket{\pi_j}_{A'_2}...\ket{\psi_k}_{A_N}\ket{\pi_j}_{A'_N}\nonumber\\
&&\otimes\ket{\phi_j}_{B_1...B_M}\nonumber\\
&&=\frac{1}{d^{(N+1)/2}}\sum_{m}\sum_{n_1,n_2,...,n_N}\ket{\Phi_{m,n_1}}\ket{\Phi_{m,n_2}}...\ket{\Phi_{m,n_N}}\nonumber\\
&\times&\sum_k\mbox{exp}\left[-\frac{2\pi ik}{d}(n_1+n_2+...+n_N)\right]\alpha_k\ket{\phi_{\ol{k+m}}}.
\eeqa

The protocol consists of three steps.

(i) Each sender performs a measurement of his particles in the generalized Bell basis
 as is shown in Fig. 1.

(ii) The senders communicate the measurement result to the $M$ receivers.
Suppose the outcome of the senders' Bell measurement is:
\beq
\ket{\Phi_{m,n_1}}\ket{\Phi_{m,n_2}}...\ket{\Phi_{m,n_N}}.
\eeq

(iii) Then, the receivers apply a local RUO that satisfies:
\beq \label{ulocal}
V_{m;n_1,n_2,...,n_N}\ket{\phi_k}=\mbox{exp}\left[\frac{2\pi ik}{d}(n_1+n_2+...+n_N)\right]\ket{\phi_{\ol{k-m}}}.
\eeq
Therefore, we have faithfully encoded the information of the initial $N$-particle state (\ref{initial}) into the $M$-particle state
\beq \label{mtm}
\ket{\phi}=\sum_{j=0}^{d-1}\alpha_j\ket{\phi_j}_{B_1B_2...B_M}.
\eeq
The second scenario is more efficient since it requires fewer entanglement
resources between the senders and receivers, and the senders perform
the measurement only $N$ times, and therefore need to transmit fewer bits of
classical information, than in the first scenario.

Although the final state is identical in the two processes one-to-many and many-to-many [see Eqs. (\ref{otm}), and (\ref{mtm})], there are some differences between them, which should be mentioned.
In the one-to-many protocol, the initial state is a multiparticle state that belongs to one sender, while in our protocol the initial multiparticle state is an entangled one shared by different observers separated in space. Also
in our scheme, the $N$ senders perform {\it locally} $N$ generalized Bell measurements. For the particular case $N=1$, we recover the result of Ref. \cite{Murao2}.
 We will show
in the last
section how one can use this generalized procedure to broadcast
entanglement to different locations.

%-----------------------------------------------------------------------------------------
\begin{figure}
\includegraphics{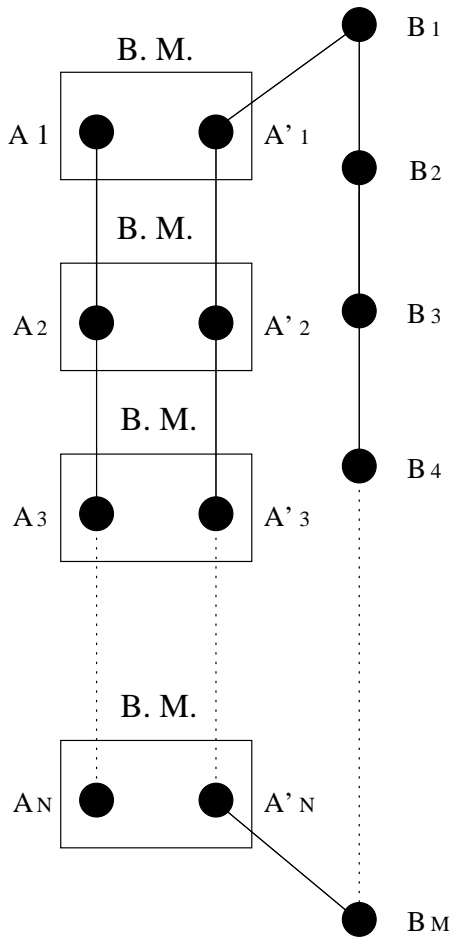} 
\vspace{0.5cm}
\caption{Schematic of the many-to-many teleportation 
 of the information of a $d$-level
system from $N$ spatially separated senders to $M$ receivers (with
$M>N$). The unknown state to be teleported is an
entangled $N$-particle state shared by $N$ observers $A_1, A_2,...A_N$.
The quantum channel
used 
 is a maximally entangled $(N+M)$-particle state.
 Each sender has to perform locally a Bell-type measurement (B.M.) and announce the
 outcome to the receivers, who perform a LRUO, obtaining the
 information of the initial unknown state. The lines represent the entanglement.
}        
       
\end{figure}

%-----------------------------------------------------------------------------------------
%----------------------------------------------------------------------------------------- 

\section{Asymmetric quantum telecloning of $d$-level systems}

\subsection{Asymmetric quantum cloning}

In this subsection we review the concept of asymmetric quantum cloning
machines for qudits ($d$-level systems) introduced in
Ref. \cite{Cerf1}. These machines produce two nonidentical clones
that emerge from two distinct Heisenberg channels.
Suppose that the input state we wish to clone $\kpsi_B $ is prepared
in the maximally entangled state $\ket{\Phi_{0,0}}$, given by
Eq. (\ref{fimn}), with a reference state, denoted by the subscript $R$.
The cloner is described by a unitary operation $U$, acting on a three-particle state, namely, the initial
state and another two $d$-level systems initially prepared in the state
$\ket{0}$, denoted by $C$ and $D$ \cite{Durt}:
\beq
U\ket{\Phi_{0,0}}_{RB}\ket{00}_{CD}=\kpsi_{RBCD}.
\eeq
 Cerf defined and analyzed the asymmetric cloning machine using the wave function of the whole system of four particles \cite{Cerf1}:
\beqa \label{wave}
\kpsi_{RBCD}&:=&\sum_{m,n=0}^{d-1}\beta_{m,n}\ket{\Phi_{m,n}}_{RB}\ket{\Phi_{m,-n}}_{CD}\nonumber\\
&=&\sum_{m,n=0}^{d-1}\gamma_{m,n}\ket{\Phi_{m,n}}_{RC}\ket{\Phi_{m,-n}}_{BD},
\eeqa 
where $B$, and $C$ denote the clones, $D$ is the ancilla, $\ket{\Phi_{m,n}}$ is given by Eq. (\ref{fimn}), and 
\beq
\gamma_{m,n}=\frac{1}{d}\sum_{x,y=0}^{d-1}e^{2\pi i(nx-my)/d}\beta_{x,y}.
\eeq
Projecting the reference on $\ket{\psi^*}$ in Eq. (\ref{wave}), we obtain the action of the cloner on an arbitrary input state \cite{Durt}:
\beq \label{clonCerf}
U\kpsi_B\ket{00}_{CD}=\sum_{m,n=0}^{d-1}\beta_{m,n}U_{m,n}\kpsi_B\ket{\Phi_{m,-n}}_{CD},
\eeq
where $\sum_{m,n=0}^{d-1}|\beta_{m,n}|^2=1$ and $U_{m,n}$ are the error operators which define the Heisenberg group \cite{Cerf1}:
\beq
U_{m,n}=\sum_{k=0}^{d-1}e^{2\pi ikn/d}\ket{\ol{k+m}}\bra{k}. 
\eeq
For $d=2$,
$U_{m,n}$ become the Pauli operators. 
In particular, for $\kpsi=\ket{j}$, Eq. (\ref{clonCerf}) becomes
\beq \label{fij}
\ket{\phi_j}:=U\ket{j}_B\ket{00}_{CD}=\sum_{m,s=0}^{d-1}b_{m,s}^{(j)}\ket{\ol{j+m}}_B\ket{s}_{C}\ket{\ol{s+m}}_D,
\eeq
 where $\{\ket{j}\}$ is the computational basis and we have defined
\beq 
b_{m,s}^{(j)}:=\frac{1}{\sqrt d}\sum_{n=0}^{d-1}\beta_{m,n}e^{2\pi i(j-s)n/d}.
\eeq
The states of Eq. (\ref{fij}) can be written in the equivalent form
\beq \label{fiequiv}
\ket{\phi_j}=\sum_{m,r=0}^{d-1}b_{m,r}\ket{\ol{j+m}}_B\ket{\ol{j+r}}_{C}\ket{\ol{j+r+m}}_D,
\eeq
with 
\beq  \label{bms} b_{m,r}:=b_{m,r}^{(0)}=\frac{1}{\sqrt d}\sum_{n=0}^{d-1}\beta_{m,n}e^{-2\pi irn/d}. \eeq 

Let us consider that we wish to clone an arbitrary $d$-level system
$\kpsi=\sum_{j=0}^{d-1}\alpha_j\ket{j}$. The state of the
three-particle system after applying the cloning machine given by
Eq. (\ref{clonCerf}) is

\beq \label{staclon}
\ket{\Pi}=U\kpsi_B\ket{0}_C\ket{0}_D=\sum_{j=0}^{d-1}\alpha_j\ket{\phi_j}.
\eeq
The two output states are described by the density operators \cite{Cerf1,Durt}
\beqa \label{output}
\rho_B&=&Tr_{CD}\proj{\Pi}{\Pi}=\sum_{m,n=0}^{d-1}|\beta_{m,n}|^2\proj{\psi_{m,n}}{\psi_{m,n}},\nonumber\\
\rho_C&=&Tr_{BD}\proj{\Pi}{\Pi}=\sum_{m,n=0}^{d-1}|\gamma_{m,n}|^2\proj{\psi_{m,n}}{\psi_{m,n}},
\eeqa
where
 \beq \label{psimn} \ket{\psi_{m,n}}=U_{m,n}\kpsi.\eeq 
The cloning machine (\ref{clonCerf}) is called a Heisenberg cloning machine, because the clones (\ref{output}) emerge from two Heisenberg channels (of probabilities $|\beta_{m,n}|^2$ and $|\gamma_{m,n}|^2$) \cite{Cerf1}.

%-----------------------------------------------------------------------------------------

\subsection{Optimal universal asymmetric Heisenberg cloning machine}

In the following, we shall analyze the class of universal asymmetric
cloning machines for $d$-level systems using a different method from the
one presented in Ref. \cite{Cerf1}. These machines generate clones that are characterized by fidelities which are independent of the input state. This condition is given by \cite{Cerf1}
\beqa |\beta_{m,n}|&=&\mu, \hspace{0.5cm} \forall \hspace{0.5cm}(m,n)\ne (0,0),\nonumber\\
|\gamma_{m,n}|&=&\eta, \hspace{0.5cm} \forall \hspace{0.5cm}(m,n)\ne (0,0).
\eeqa

The most interesting case is when the universal cloning machine is
optimal, which means a machine that creates the second clone with
maximal fidelity for a given fidelity of the first one
\cite{Cerf1,Durt}. In Ref. \cite{Cerf1} it was shown that the optimal
machines are characterized by $\beta_{m,n} $, $\gamma_{m,n} $ that
have the same phases; therefore it is sufficient to analyze the case
when $\beta_{m,n}$ and $\gamma_{m,n}$ are real.
With the notation $\beta_{0,0}=\nu $, we get from Eq. (\ref{bms})
\beqa
b_{0,0}&=&\frac{1}{\sqrt d}[\nu+(d-1)\mu],\nonumber\\
b_{m,0}&=&\sqrt d\mu,\nonumber\\
b_{0,r}&=&\frac{1}{\sqrt d}(\nu-\mu),\nonumber\\
b_{m,r}&=&0, \hspace{0.5cm} (m,r)\ne (0,0).
\eeqa
Therefore, we find the states (\ref{fiequiv}) that define the cloning machine as
\beqa
&&\ket{\phi_j}=U\ket{j}\ket{00}=\frac{1}{\sqrt{\cal R}}(\frac{1}{\sqrt d}[\nu+(d-1)\mu]\ket{j}\ket{j}\ket{j}\nonumber\\
&+&\frac{1}{\sqrt d}(\nu-\mu)\sum_{r=1}^{d-1}\ket{j}\ket{\ol{j+r}}\ket{\ol{j+r}}+\sqrt d\mu\sum_{r=1}^{d-1}\ket{\ol{j+r}}\ket{j}\ket{\ol{j+r}}), 
\eeqa
where ${\cal R}$ is the normalization factor. 
We obtain the general expression for the optimal universal asymmetric Heisenberg cloning machine:
\beqa \label{general}
&&U\ket{j}\ket{00}
=\frac{1}{\sqrt{1+(d-1)(p^2+q^2)}}( \ket{j}\ket{j}\ket{j}+p\sum_{r=1}^{d-1}\ket{j}\ket{\ol{j+r}}\ket{\ol{j+r}}\nonumber\\
&&+q\sum_{r=1}^{d-1}\ket{\ol{j+r}}\ket{j}\ket{\ol{j+r}}) , 
\eeqa
where $p=(\nu-\mu)/[\nu+(d-1)\mu]$ and $q=d\mu/[\nu+(d-1)\mu]=1-p$. 

Suppose we wish to clone the state $\kpsi=\sum_{j=0}^{d-1}\alpha_j\ket{j}$. The total state of the two clones and ancilla (after cloning) is given by Eq. (\ref{staclon}):
\beqa
\ket{\Pi}&=&\frac{1}{\sqrt{1+(d-1)(p^2+q^2)}}\sum_{j=0}^{d-1}\alpha_j(
  \ket{j}\ket{j}\ket{j}+p\sum_{r=1}^{d-1}\ket{j}\ket{\ol{j+r}}\ket{\ol{j+r}}\nonumber\\
&+&q\sum_{r=1}^{d-1}\ket{\ol{j+r}}\ket{j}\ket{\ol{j+r}}), 
\eeqa
where $p+q=1$.
We find the state of the clones:
\beq
\rho_B=\frac{1}{1+(d-1)(p^2+q^2)}\left\{ \left[ 1-q^2+(d-1)p^2\right] \proj{\psi}{\psi}+q^2I\right\}
\eeq
and
\beq
\rho_C=\frac{1}{1+(d-1)(p^2+q^2)}\left\{ \left[ 1-p^2+(d-1)q^2\right] \proj{\psi}{\psi}+p^2I\right\}.
\eeq

The fidelity of two mixed states is defined as \cite{Jozsa}
\beq \label{fidelitate}
F(\rho_1,\rho_2)=\left[ \mbox{Tr}\left( \sqrt{\rho_1}\rho_2\sqrt{\rho_1}\right)^{1/2}\right]^2.
\eeq
In the particular case when one of the two states is pure,
Eq. (\ref{fidelitate}) becomes
\beq 
F(\proj{\psi}{\psi},\rho)=\bra{\psi}\rho \ket{\psi}.
\eeq
 We get the fidelities of the two clones as
\beq
F_B(\proj{\psi}{\psi} , \rho_B)=\frac{1+(d-1)p^2}{1+(d-1)\left( p^2+q^2\right)}
\eeq
and
\beq
F_C(\proj{\psi}{\psi} ,\rho_C)=\frac{1+(d-1)q^2}{1+(d-1)\left( p^2+q^2\right)}.
\eeq
In Ref. \cite{Werner}, Werner investigated the $N\to M$ universal symmetric cloning machine for $d$-dimensional systems. He found the fidelity of the optimal cloner as
\beq \label{FW}
F=\frac{N(d+M-1)+M}{M(d+N)}.
\eeq
The symmetric cloning machine is obtained when $p=q=1/2$ in Eq. (\ref{general}), and this leads to the fidelity of the two clones
\beq
F_B=F_C=\frac{d+3}{2(d+1)}
\eeq 
which is in agreement with Eq. (\ref{FW}).

The Heisenberg cloning machine generates the best copies when
\beq
F_B+F_C=1+\frac{1}{1+(d-1)[p^2+(1-p)^2]}
\eeq
is maximal. In order to evaluate the maximum, we calculate
\beq
\frac{\partial (F_B+F_C)}{\partial p}=\frac{-2(d-1)(2p-1)}{\{1+(d-1)[p^2+(1-p)^2]\}^2}
\eeq
and 
\beq
\frac{\partial^2 (F_B+F_C)}{\partial p^2}|_{p=1/2}=-16\frac{d-1}{(d+1)^2}<0.
\eeq
We see that we recover the result of Ref. \cite{Cerf1}, namely, 
 the universal symmetric quantum cloning machine ($p=1/2$) will
optimally copy the state. The general expression for the optimal
universal asymmetric Heisenberg cloning machine (\ref{general})
obtained here is more useful when we investigate the telecloning
process and broadcasting of entanglement than the description that contains the reference system.

%-----------------------------------------------------------------------------------------

\subsection{Asymmetric quantum telecloning of $d$-level systems}

In this subsection we extend the concept of asymmetric telecloning from qubits to $d$-dimensional systems. Therefore, we 
shall study the $1\to 2$ asymmetric quantum
telecloning of qudits, a protocol that combines quantum teleportation and asymmetric cloning from one observer to two parties, where the outputs are produced by Cerf's Heisenberg cloning machine.
Consider that Alice holds an unknown $d$-level system
\beq \label{partA}
\kpsi_A=\sum_{k=0}^{d-1}\alpha_k\ket{k}_A.
\eeq
 She wishes to distribute the information of this $d$-level system to two distant receivers Bob and Charlie with different fidelities. Since Alice cannot produce perfect copies of an unknown quantum state due to the no-cloning theorem \cite{Wootters}, the best way to transmit the information is to  send two optimal asymmetric clones to Bob and Charlie.

We define the channel used in the telecloning process as a four-particle state shared by Alice, Bob, Charlie, and a fourth person Daniel:
\beq \label{channel}
\ket{\xi}_{ABCD}=\frac{1}{\sqrt d}\sum_{j=0}^{d-1}\ket{j}_A\ket{\phi_j}_{BCD},
\eeq 
where $A$, $B$, $C$, and $D$ denote the particles that belong to Alice, Bob, Charlie, and Daniel, respectively. The states $\ket{\phi_j}_{BCD}$ are given by Eq. (\ref{fiequiv}). The tensor product between Alice's input state of Eq. (\ref{partA}) and the quantum channel is
\beq
\kpsi_A\otimes \ket{\xi}_{ABCD}=\sum_{n,m=0}^{d-1}\ket{\Phi_{mn}}_A\otimes \frac{1}{d}\sum_{j=0}^{d-1}e^{-2\pi ijn/d}\alpha_j\ket{\phi_{\ol{j+m}}}_{BCD}.
\eeq

We shall apply the one-to-many teleportation protocol presented in a
previous section in order to teleclone the initial state. Therefore we have
to find the LRUO that has to be performed by Bob,
Charlie, and Daniel in order to recover the information of Alice's
$d$-level system. 
Let us define the LRUO using the same operators given in
Ref. \cite{Murao2}, which were introduced in order to perform $1\to M
$ symmetric telecloning of a $d$-level system:
\beq 
V_{m;n}={\cal V}_{mn}^B\otimes {\cal V}_{mn}^C\otimes {\cal V}_{mn}^D,
\eeq 
where
\beq \label{op1}
{\cal V}_{mn}^X=\sum_{j=0}^{d-1}e^{2\pi ijn/d}\ket{j}\bra{\ol{j+m}}, \hspace{0.5cm} X=B,C,
\eeq
and
\beq \label{op2}
{\cal V}_{mn}^D=\sum_{j=0}^{d-1}e^{-2\pi ijn/d}\ket{j}\bra{\ol{j+m}}.
\eeq

In Ref. \cite{Murao2} the asymmetric telecloning of two-level states was investigated, where the output states are generated by the optimal universal asymmetric Pauli cloning machine [given by Eq. (\ref{general}) for $d=2$]. It has been shown that the LRUO is given by Eqs. (\ref{op1}) and (\ref{op2}) by using a symmetry property of the quantum channel under the permutations of particles that belong to two groups of observers (Alice and Daniel) and (Bob and Charlie). 
Although this symmetry condition is not satisfied by the channel
(\ref{channel}), we shall prove that the LRUO introduced in
Ref.\cite{Murao2} is also useful for the asymmetric telecloning of an
arbitrary $d$-level system.
All we need to show is 
that $V_{m;n}$ satisfies Eq. (\ref{onetomany}):
\beq
V_{m;n}\ket{\phi_j}=\sum_{m,r=0}^{d-1}b_{m,r}e^{2\pi ijn/d}\ket{j}\ket{\ol{j+r-m}}\ket{\ol{j+r}}=e^{2\pi ijn/d}\ket{\phi_{\ol{j-m}}},
\eeq
and
therefore Bob, Charlie, and Daniel will share the state of Eq. (\ref{staclon})
\beq
 \ket{\Pi}=\sum_{j=0}^{d-1}\alpha_j\ket{\phi_j}_{BCD}.
\eeq
Then, following the description presented in Sec. III A, Bob and Charlie can obtain the Heisenberg asymmetric clones
(\ref{output})
\beqa 
\rho_B&=&\sum_{m,n=0}^{d-1}|\beta_{m,n}|^2\proj{\psi_{m,n}}{\psi_{m,n}}\nonumber\\
\rho_C&=&\sum_{m,n=0}^{d-1}|\gamma_{m,n}|^2\proj{\psi_{m,n}}{\psi_{m,n}}.
\eeqa
Therefore, the information contained in the initial $d$-dimensional
system has been copied and transmitted at the same time to two receivers with different
fidelities, using quantum telecloning.

%-----------------------------------------------------------------------------------------
%-----------------------------------------------------------------------------------------

\section{Broadcasting of inseparability using optimal
  universal asymmetric cloning machines}
\subsection{Broadcasting of entanglement using local optimal
  universal asymmetric cloners}

Following the work of  Ref. \cite{Buzek2}, we investigate the broadcasting of entanglement using a more general cloner, namely,
the optimal universal asymmetric Pauli cloning machine.
 The
 question we address now is the following: How well can the observers copy an entangled state by
 applying local asymmetric cloning transformations? We shall analyze
 the efficiency of the broadcasting process by calculating the fidelities of the two output states.

In Ref. \cite{Buzek2},  Bu$\check z$ek {\it et al.} proposed the
broadcasting of entanglement using local optimal universal
symmetric cloning machines defined as \cite{Bruss}
\beqa 
U\ket{0}\ket{00}=\sqrt{\frac{2}{3}}\left( \ket{000}+\frac{1}{2}\ket{011}+\frac{1}{2}\ket{101}\right),\nonumber\\
U\ket{1}\ket{00}=\sqrt{\frac{2}{3}}\left( \ket{111}+\frac{1}{2}\ket{100}+\frac{1}{2}\ket{010}\right).
\eeqa
The first two qubits represent the clones and the last
one is the ancilla.

Let us now investigate the case when two distant observers (Alice and Bob) apply locally an asymmetric cloning machine. The initial entanglement shared by Alice and Bob is
\beq \label{entini}
\kpsi_{12}=\alpha \ket{00}+\beta \ket{11}.
\eeq
 We assume  for
simplicity that $\alpha$ 
 and  $\beta$ are real.
Alice and Bob use the optimal universal asymmetric Pauli cloning machine given by Eq.(\ref{general}) for $d=2$:
\beqa \label{Pauli}
U\ket{0}\ket{00}=\frac{1}{\sqrt{1+p^2+q^2}}(\ket{000}+p\ket{011}+q\ket{101}),\nonumber\\
U\ket{1}\ket{00}=\frac{1}{\sqrt{1+p^2+q^2}}(\ket{111}+p\ket{100}+q\ket{010}),
\eeqa
with $p+q=1$. 
Therefore, the state of the total system, consisting of the two particles 1 and 2, and another four particles, the blank states 3 and 4, and the ancillas 5, and 6, after applying the cloning transformation (\ref{Pauli}) by Alice and Bob, is
\beqa \label{pi'}
&&\ket{\Pi'}=U\otimes U\kpsi_{12}\ket{00}_{35}\ket{00}_{46}=\frac{1}{\sqrt{1+p^2+q^2}}\{\ket{00}_{56}[\alpha \ket{00}_{13}\ket{00}_{24}\nonumber\\
&&+\beta p^2\ket{10}_{13}\ket{10}_{24}
+\beta pq\ket{10}_{13}\ket{01}_{24}+\beta pq\ket{01}_{13}\ket{10}_{24}+\beta q^2\ket{01}_{13}\ket{01}_{24}]\nonumber\\
&&+\ket{01}_{56}[\alpha p\ket{00}_{13}\ket{01}_{24}+\alpha q\ket{00}_{13}\ket{10}_{24}+\beta p\ket{10}_{13}\ket{11}_{24}+\beta q\ket{01}_{13}\ket{11}_{24}]\nonumber\\
&&+\ket{10}_{56}[\alpha q\ket{10}_{13}\ket{00}_{24}+\alpha p\ket{01}_{13}\ket{00}_{24}+\beta p\ket{11}_{13}\ket{10}_{24}+\beta q\ket{11}_{13}\ket{01}_{24}]\nonumber\\
&&+\ket{11}_{56}[\alpha p^2\ket{01}_{13}\ket{01}_{24}+\alpha pq\ket{01}_{13}\ket{10}_{24}\nonumber\\
&&+\alpha pq\ket{10}_{13}\ket{01}_{24}+\alpha
q^2\ket{10}_{13}\ket{10}_{24}+\beta
\ket{11}_{13}\ket{11}_{24}]\}\nonumber\\
&&=\alpha \ket{\phi_0}+\beta \ket{\phi_1},
\eeqa
where the particles denoted by odd numbers belong to Alice, while the even particles belong to Bob. 

We say that the input state $\kpsi_{12}$ has been broadcast if the
following two necessary conditions are satisfied \cite{Buzek2}:
(i) the local reduced density operators $\rho_{13}$ and $\rho_{24}$ are separable, and 
(ii) the nonlocal states $\rho_{14}$ and $\rho_{23}$ are inseparable.

 Let us recall the definition of the inseparability of a bipartite system: A state is inseparable if the density operator describing this state cannot be written as a convex combination of product states \cite{Werner2}
\beq
\rho=\sum_ip_i\rho_i^{(1)}\otimes \rho_i^{(2)}.
\eeq 
We shall explicitly rewrite the above relation:
\beq
\rho_{m\mu,n\nu}=\sum_ip_i\left(\rho_i^{(1)}\right)_{mn}\otimes \left(\rho_i^{(2)}\right)_{\mu \nu}.
\eeq 
Let us define a matrix $\sigma $, which is the partial
transposition of $\rho $:
\beq
\sigma_{m\mu,n\nu}:=\rho_{n\mu,m\nu}.
\eeq
The necessary condition for separability of a bipartite state is given
by Peres' statement, that the eigenvalues of the partial
transpose of the density matrix must be positive
\cite{Peres}. Horodecki {\it et al.} have shown that this condition is also
sufficient for the separability of a mixed spin state of two spin $1/2$ particles \cite{Horodecki}.

We find the expression of the reduced density operators of the local states:
\beqa
\rho_{13}=\rho_{24}&=&\frac{1}{(1+p^2+q^2)^2}[\alpha^2(1+p^2+q^2)\proj{00}{00}\
+\beta^2(1+p^2+q^2)\proj{11}{11}\nonumber\\
&&+(p^2q^2+\beta^2q^4+\beta^2q^2+\alpha^2p^4+\alpha^2p^2)\proj{01}{01}\nonumber\\
&&+(p^2q^2+\beta^2p^4+\beta^2p^2+\alpha^2q^4+\alpha^2q^2)\proj{10}{10}\nonumber\\
&&+(pq+p^3q+pq^3)(\proj{01}{10}+\proj{10}{01})].
\eeqa
Applying the Peres-Horodecki theorem, we get the condition for the separability of the local states
\beq
\alpha^2\beta^2-p^2q^2\ge 0
\eeq
or equivalently
\beq
\frac{1}{2}\left[ 1-\sqrt{1-4p^2(1-p)^2}\right]\le \alpha^2 \le \frac{1}{2}\left[ 1+\sqrt{1-4p^2(1-p)^2}\right].
\eeq

The nonlocal pairs of particles are described by the following density operators:
\beqa \label{nonl1}
\rho_{14}&=&\frac{1}{(1+p^2+q^2)^2}\{[p^2q^2+\alpha^2(1+p^2+q^2)]\proj{00}{00}+[p^2q^2\nonumber\\
&&+\beta^2(1+p^2+q^2)]\proj{11}{11}+4pq\alpha\beta(\proj{00}{11}+\proj{11}{00})\nonumber\\
&&+(\beta^2q^4+\beta^2q^2+\alpha^2p^4+\alpha^2p^2)\proj{01}{01}\nonumber\\
&&+(\beta^2p^4+\beta^2p^2+\alpha^2q^4+\alpha^2q^2)\proj{10}{10}\}
\eeqa
and
\beqa \label{nonl2}
\rho_{23}&=&\frac{1}{(1+p^2+q^2)^2}\{[p^2q^2+\alpha^2(1+p^2+q^2)]\proj{00}{00}+[p^2q^2\nonumber\\
&&+\beta^2(1+p^2+q^2)]\proj{11}{11}+4pq\alpha\beta(\proj{00}{11}+\proj{11}{00})\nonumber\\
&&+(\beta^2p^4+\beta^2p^2+\alpha^2q^4+\alpha^2q^2)\proj{01}{01}\nonumber\\
&&+(\beta^2q^4+\beta^2q^2+\alpha^2p^4+\alpha^2p^2)\proj{10}{10}\}.
\eeqa
Then the two nonlocal states are inseparable if
\beq
(\beta^2p^4+\beta^2p^2+\alpha^2q^4+\alpha^2q^2)(\beta^2q^4+\beta^2q^2+\alpha^2p^4+\alpha^2p^2)-16\alpha^2\beta^2p^2q^2\le 0
\eeq
or equivalently
\beq \label{alfa1}
\frac{1}{2}\left( 1-\sqrt{1-4\lambda }\right)\le \alpha^2 \le \frac{1}{2}\left( 1+\sqrt{1-4\lambda }\right),
\eeq
where
 \beq \lambda=\frac{p^4q^4+p^2q^4+p^4q^2+p^2q^2}{2p^4q^4+2p^4q^2+2p^2q^4-q^8-2q^6-q^4-p^8-2p^6-p^4+18p^2q^2}.\eeq
The requirements that $1-4\lambda$ has to be positive and that
 the local states are separable when the nonlocal ones are inseparable lead to
\beq \label{condp}
\frac{1}{2}\left(1-\sqrt{-9+2\sqrt{21}}\right)\le p\le \frac{1}{2}\left(1+\sqrt{-9+2\sqrt{21}}\right)
.
\eeq
The symmetric cloning machine $(p=1/2)$ discussed in Ref. \cite{Buzek2} satisfies the condition (\ref{condp}).
We see from Eq. (\ref{alfa1}) that only some entangled states can be copied. 

The fidelities of the nonlocal states are given by
\beq \label{fidloc}
F(\kpsi_{12},\rho_{14})=F(\kpsi_{12},\rho_{23})=\frac{p^2q^2}{(1+p^2+q^2)^2}+\frac{\alpha^4+\beta^4}{1+p^2+q^2}+\frac{8\alpha^2\beta^2pq}{(1+p^2+q^2)^2}.
\eeq
In order to find the maximal value of the fidelity we compute
\beqa
\frac{\partial F}{\partial p}&=&\frac{2p-1}{[1+p^2+(1-p)^2]^4}[8p(1-p+p^2)-8(1-p+p^2)^2(\alpha ^4+\beta ^4)\nonumber\\
&-&32(1-p+p^2)(1+2p-2p^2)\alpha ^2\beta ^2].
\eeqa
Since we have
\beq
\frac{\partial ^2F}{\partial p^2}|_{p=1/2}=-\frac{16}{27}(1+10 \alpha ^2\beta ^2)<0,
\eeq
the maximal fidelity is obtained for $p=1/2$, which means that the local symmetric
cloning machine generates optimally the two final entangled states. Bandyopadhyay and Kar have analyzed the broadcasting of entanglement by using the most
general local symmetric cloning machines, and also found that the optimal universal symmetric machines give the best result \cite{SB}. Therefore, we conclude that the universal symmetric machine will optimally accomplish the task. 

>From Eqs. (\ref{nonl1}) and (\ref{nonl2}), we see that the two output entangled
states cannot be writen, in general, in the scaled form
\beqa \label{impclon}
\rho_{14}&=&\eta_1\proj{\psi }{\psi }+(1-\eta_1 )\frac{1}{4}I\otimes
I\nonumber\\
\rho_{23}&=&\eta_2\proj{\psi }{\psi }+(1-\eta_2 )\frac{1}{4}I\otimes
I.
\eeqa
Only for $\alpha =\beta =1/\sqrt 2$ are the two nonlocal states
in the scaled form (\ref{impclon}) with
\beq
\eta_1=\eta_2=\frac{pq}{(1-pq)^2}.
\eeq
Therefore, by applying local asymmetric cloning machines on the
maximally entangled state we obtain
a universal symmetric cloning machine of entanglement. This
generalizes the result of Ref. \cite{Buzek3}, where local symmetric
cloning machines were used.

%-----------------------------------------------------------------------------------------

\subsection{Nonlocal asymmetric cloning of an entangled state}

Bu$\check z$ek {\it et al.} showed that the entanglement is better
copied when a nonlocal symmetric machine is applied instead of two local
symmetric cloners \cite{Buzek3}.
In this subsection we shall investigate cloning of entanglement considering optimal universal {\it asymmetric} machines and then 
compare the two methods (local and
nonlocal).

Let us apply the nonlocal optimal cloning machine of
Eq. (\ref{general}) for $d=4$ on the
state
\beq
 \kpsi=\alpha \ket{00}+\beta \ket{11}.
\eeq
We obtain the two density operators that describe the clones as
\beq
\rho_1=\frac{1}{1+3(p^2+q^2)}\left[ \left( 1-q^2+3p^2\right) \proj{\psi}{\psi}+q^2I\right]
\eeq
and
\beq
\rho_2=\frac{1}{1+3(p^2+q^2)}\left[ \left( 1-p^2+3q^2\right) \proj{\psi}{\psi}+p^2I\right].
\eeq
From the Peres-Horodecki theorem it follows that $\rho_j$ 
is inseparable if
\beq \label{alfa}
\frac{1}{2}\left( 1-\sqrt{1-4\mu_j }\right)\le \alpha^2 \le \frac{1}{2}\left( 1+\sqrt{1-4\mu_j }\right),
\eeq
where $j=1, 2$, and
\beqa
 \mu_1&=&\frac{(1-p)^4}{4p^2(p+1)^2},\nonumber\\
\mu_2&=&\frac{(1-q)^4}{4q^2(q+1)^2}.
\eeqa
Imposing the condition that $1-4\mu_j$ has to be positive for $j=1,2$,
we find
\beq \label{pcond}
\frac{1}{3}\le p\le \frac{2}{3}.
\eeq
Combining the conditions (\ref{alfa}) and (\ref{pcond}) we obtain 
 (i) if $p\in \left[ \frac{1}{3},\frac{1}{2}\right)$ then
\beq \label{interval1}
\frac{1}{2}\left( 1-\sqrt{1-4\mu_2 }\right)\le \alpha^2 \le \frac{1}{2}\left( 1+\sqrt{1-4\mu_2 }\right);
\eeq
and (ii) if $p\in \left[ \frac{1}{2},\frac{2}{3}\right]$ then
\beq \label{interval2}
\frac{1}{2}\left( 1-\sqrt{1-4\mu_1 }\right)\le \alpha^2 \le \frac{1}{2}\left( 1+\sqrt{1-4\mu_1 }\right).
\eeq
The fidelities of the two output states are
\beqa
F_B(\proj{\psi}{\psi} , \rho_B)&=&\frac{1+3 p^2}{1+3\left( p^2+q^2\right)},\nonumber\\
F_C(\proj{\psi}{\psi} ,\rho_C)&=&\frac{1+3 q^2}{1+3\left( p^2+q^2\right)}.
\eeqa
We showed in Sec. IV A that the two output states produced using local
asymmetric cloning are inseparable for states that satisfy Eq. (\ref{alfa1}).

If one compares the range of $\alpha^2$ obtained using local and
nonlocal cloners, one observes that
 the range of $\alpha^2$ given by Eqs. (\ref{interval1}) and
(\ref{interval2}) is wider than the range of Eq. (\ref{alfa1}); therefore we
conclude that the entanglement is better copied using a nonlocal
asymmetric cloner than two local asymmetric cloners. This is a
generalization of the result of Ref. \cite{Buzek3}, where symmetric
cloning machines ($p=1/2$) were considered.

%-----------------------------------------------------------------------------------------
%-----------------------------------------------------------------------------------------

\section{Telebroadcasting of a bipartite two-level entangled state }

Now we shall apply the many-to-many teleportation protocol described in
Sec. II for $1 \to2$ {\it telebroadcasting}, a process that combines
teleportation and broadcasting, of an arbitrary bipartite two-level entangled state.  
Let us investigate the following scenario. Assume that two spatially separated
observers $A_1$ and $A_2$ hold an entangled state and they wish to send two copies of
this state to two pairs of observers also located at different
places. A simple way for $A_1$ and $A_2$ to accomplish the task is, first,
to broadcast the entanglement using local cloning machines, as we have
shown in the previous section, and, second, to teleport each clone
using the standard procedure \cite{Bennett}. Here we propose a
scheme that is more efficient in terms of the entanglement shared
between senders and receivers, and classical communication.

Suppose two observers $A_1$ and $A_2$ share an entangled state
\beq
 \kpsi_{A_1A_2}=\alpha \ket{00}+\beta \ket{11}
\eeq
 that they wish to teleport to two pairs of receivers $B_1$, $B_4$, and
 $B_2$, $B_3$, respectively. 
We shall construct the quantum channel using the six-particle states,
which are specified in Eq. (\ref{pi'}),
\beqa
\ket{\phi_0}_{B_1B_2B_3B_4B_5B_6}&=&\frac{1}{1+p^2+q^2}(\ket{000000}+p\ket{000101}+q\ket{010001}\nonumber\\
&&+p\ket{001010}+p^2\ket{001111}+pq\ket{011011}+q\ket{100010}\nonumber\\
&&+pq\ket{100111}+q^2\ket{110011}),
\eeqa
\beqa
\ket{\phi_1}_{B_1B_2B_3B_4B_5B_6}&=&\frac{1}{1+p^2+q^2}(\ket{111111}+p\ket{111010}+q\ket{101110}\nonumber\\
&&+p\ket{110101}+p^2\ket{110000}+pq\ket{100100}+q\ket{011101}\nonumber\\
&&+pq\ket{011000}+q^2\ket{001100}),
\eeqa
where $B_5$ and $B_6$ are the two observers who hold the ancillas.
We choose the quantum channel as
\beq \label{telecstate}
\ket{\xi}=\frac{1}{\sqrt 2}\ket{00}_{A'_1A'_2}\ket{\phi_0}_{B_1B_2B_3B_4B_5B_6}+\frac{1}{\sqrt 2}\ket{11}_{A'_1A'_2}\ket{\phi_1}_{B_1B_2B_3B_4B_5B_6}.
\eeq

Let us define the LRUO $V_{m;n_1,n_2}$ which satisfies Eq. (\ref{ulocal}) as
\beqa \label{ubip}
&V_{0;0,0}=V_{0;1,1}&=I,\nonumber\\
&V_{0;0,1}=V_{0;1,0}&=\sigma_z\otimes I\otimes \sigma_z\otimes I \otimes \sigma_z\otimes I,\nonumber\\
&V_{1;0,0}=V_{1;1,1}&=\sigma_x\otimes \sigma_x\otimes \sigma_x\otimes \sigma_x\otimes \sigma_x\otimes \sigma_x, \nonumber\\
&V_{1;1,0}=V_{1;0,1}&=\sigma_x \sigma_z\otimes I\otimes \sigma_x \sigma_z\otimes I\otimes\sigma_x \sigma_z\otimes I. 
\eeqa
Then, using the many-to-many protocol we get
the final state shared by the six receivers:
\beq
\ket{\Pi'}=\alpha \ket{\phi_0}_{B_1B_2B_3B_4B_5B_6}+\beta \ket{\phi_1}_{B_1B_2B_3B_4B_5B_6},
\eeq
which is identical with the state of Eq. (\ref{pi'}). Following the
procedure described in Sec. IV A, we obtain two entangled pairs
shared by two pairs of receivers $B_1-B_4$ and $B_2-B_3$,
characterized by the density operators of Eqs. (\ref{nonl1}) and (\ref{nonl2}):
\beqa
\rho_{B_1B_4}&=&\frac{1}{(1+p^2+q^2)^2}\{[p^2q^2+\alpha^2(1+p^2+q^2)]\proj{00}{00}+[p^2q^2\nonumber\\
&&+\beta^2(1+p^2+q^2)]\proj{11}{11}+4pq\alpha\beta(\proj{00}{11}+\proj{11}{00})\nonumber\\
&&+(\beta^2q^4+\beta^2q^2+\alpha^2p^4+\alpha^2p^2)\proj{01}{01}\nonumber\\
&&+(\beta^2p^4+\beta^2p^2+\alpha^2q^4+\alpha^2q^2)\proj{10}{10}\} \nonumber
\eeqa
and
\beqa
\rho_{B_2B_3}&=&\frac{1}{(1+p^2+q^2)^2}\{[p^2q^2+\alpha^2(1+p^2+q^2)]\proj{00}{00}+[p^2q^2\nonumber\\
&&+\beta^2(1+p^2+q^2)]\proj{11}{11}+4pq\alpha\beta(\proj{00}{11}+\proj{11}{00})\nonumber\\
&&+(\beta^2p^4+\beta^2p^2+\alpha^2q^4+\alpha^2q^2)\proj{01}{01}\nonumber\\
&&+(\beta^2q^4+\beta^2q^2+\alpha^2p^4+\alpha^2p^2)\proj{10}{10}\}. \nonumber
\eeqa 
 for p that
satisfies Eq. (\ref{condp}).

In conclusion, we have proved how one can transmit optimal
information of an entangled state to two pairs of receivers using only
local operations and classical communication. The final states
obtained by the four receivers represent the output systems generated
in broadcasting of the entanglement process using the local optimal universal
asymmetric cloning machines described in Sec. IV A.

%-----------------------------------------------------------------------------------------
%-----------------------------------------------------------------------------------------

\section{Conclusions and open questions}
We have proposed a generalization of quantum teleportation, called the
many-to-many protocol, that transmits the information of a $d$-level
system from $N$ spatially separated senders to $M>N$ receivers situated
at different locations. In this scheme, each sender has to perform a
Bell-type measurement and annouce the outcome to the
receivers. Depending on the result, the receivers apply a LRUO
in order to obtain the information of the initial $d$-level system. We
have shown that this protocol is more efficient than another method,
where the senders use local global operations on the $N$ particles and
$M-N$ ancillas. 
We have presented a generalization of broadcasting of entanglement
using local optimal universal asymmetric cloners and have shown that
entanglement is best copied when symmetric cloners are applied. We
have also discussed nonlocal asymmetric cloning of an entangled state
and pointed out that the inseparability is copied better using
a nonlocal asymmetric cloning machine than local asymmetric cloning
machines.
Using the one-to-many communication protocol, we have extended asymmetric telecloning from qubits to
$d$-dimensional systems. 
Finally, we have presented a scheme called telebroadcasting that
combines the many-to-many teleportation protocol and asymmetric
broadcasting of entanglement. The quantum channel is a maximally
entangled state shared between the senders and receivers. We have
proved that using Bell measurement, classical communication, and local
operations, one can
 distribute entanglement from two observers to two pairs of
spatially separated observers.

In closing, let us point out some open questions. First, is the
 telebroadcasting presented in the last section the optimal protocol
 that copies and distributes entanglement to different locations?
 Second, is it possible
 to use fewer entanglement resources than the channel of Eq. (\ref{telecstate})
 between the senders and receivers? An interesting problem would be to
 consider symmetric telebroadcasting, where the two output
 entangled states have to be generated by universal
symmetric cloning machines (not necessary optimal).

\section*{Acknowledgments}
 I wish to thank Professor Anders Karlsson and  Professor
 Gunnar Bj\"ork for useful discussions
 and critical comments on the manuscript. This work was partially supported by a
 grant from the Swedish Foundation for Strategic Research and the
 European Community through the IST FET QuComm project.

\vspace{0.3cm}

%\begin{references}

%\end{references}

\end{document}